\documentclass[twocolumn,numberedappendix,appendixfloats,iop,apjl]{emulateapj}
\usepackage{graphicx}
\usepackage{epstopdf}
\usepackage{mathrsfs}
\usepackage{amssymb}
\usepackage[title]{appendix}
\usepackage{amsmath}
\usepackage{color}
\usepackage{bm}

\newcommand{\pri}{\prime}

\newcommand{\vv}{\mathbf v}
\newcommand{\vrr}{\mathbf r}

\newcommand{\vk}{\mathbf k}
\newcommand{\vq}{\mathbf q}

\newcommand{\mH}{\mathcal H}

\newcommand{\drho}{\delta_{\rho}}

\newcommand{\vPsi}{\bm \Psi}

\newcommand{\vxi}{\bm \xi}

\newcommand{\vx}{\bm x}

\newcommand{\Mpc}{\rm Mpc}

\def\presuper#1#2%
  {\mathop{}%
   \mathopen{\vphantom{#2}}^{#1}%
   \kern-\scriptspace%
   #2}

\begin{document}

\title{Isobaric Reconstruction of the Baryonic Acoustic Oscillation}

\author{Xin Wang\altaffilmark{1,10}, 
Hao-Ran Yu\altaffilmark{1,2}, 
Hong-Ming Zhu\altaffilmark{3,4}, 
Yu Yu\altaffilmark{5}, 
Qiaoyin Pan\altaffilmark{9}, 
Ue-Li Pen\altaffilmark{1,6,7,8}}
\altaffiltext{1}{Canadian Institute for Theoretical Astrophysics, 60 St. George St., Toronto, ON,
M5H 3H8, Canada}
\altaffiltext{2}{Kavli Institute for Astronomy and Astrophysics, Peking University, Beijing 100871, China}
\altaffiltext{3}{Key Laboratory for Computational Astrophysics, National Astronomical Observatories, 
Chinese Academy of Sciences, 20A Datun Road, Beijing 100012, China}
\altaffiltext{4}{University of Chinese Academy of Sciences, Beijing 100049, China}
\altaffiltext{5}{Key Laboratory for Research in Galaxies and Cosmology, Shanghai Astronomical Observatory, 
Chinese Academy of Sciences, 80 Nandan Road, Shanghai 200030, China}
\altaffiltext{6}{Dunlap Institute for Astronomy and Astrophysics, University of Toronto,
50 St. George Street, Toronto, Ontario M5S 3H4, Canada}
\altaffiltext{7}{Canadian Institute for Advanced Research, CIFAR Program in Gravitation and Cosmology, 
Toronto, Ontario M5G 1Z8, Canada}
\altaffiltext{8}{Perimeter Institute for Theoretical Physics, 31 Caroline Street North, Waterloo, Ontario, N2L 2Y5, 
Canada 8Center of High Energy Physics, Peking University, Beijing 100871, China}
\altaffiltext{9}{School of Physics, Nankai University, 94 Weijin Rd, Nankai, Tianjin, 300071, China}
\altaffiltext{10}{xwang@cita.utoronto.ca}

\label{firstpage}

\begin{abstract}
In this paper, we report a significant recovery of the linear baryonic acoustic oscillation (BAO) signature
by applying the isobaric reconstruction algorithm to the non-linear matter density field. 
Assuming only the longitudinal component of the displacement being cosmologically relevant, this algorithm 
iteratively solves the coordinate  transform  between the Lagrangian and Eulerian frames without 
requiring any specific knowledge of the dynamics. 
For dark matter field, it produces the non-linear displacement potential with very high fidelity. 
The reconstruction error at the pixel level is within a few percent, and is caused only by the emergence of the 
transverse component after the shell-crossing.
As it circumvents the strongest non-linearity of the density evolution, the reconstructed field is 
well-described by linear theory and immune from the bulk-flow smearing of the BAO signature. 
Therefore this algorithm could significantly improve the measurement accuracy of the sound horizon scale $s$. 
For a perfect large-scale structure survey at redshift zero without Poisson or instrumental noise, the 
fractional error $\Delta s/s$ is reduced by a factor of $\sim 2.7$,  very close to the ideal limit 
with linear power spectrum and Gaussian covariance matrix. 
\end{abstract}

\keywords{large-scale structure of universe}

\maketitle

\section{Introduction}

Large-scale structure (LSS) surveys provide a wealth of cosmological information about our Universe. 
Particularly, the wiggling feature of the baryonic acoustic oscillation imprinted on the matter power spectrum 
encodes the characteristic sound horizon scale of $s \sim 100 \Mpc/h$, which serves as a standard ruler 
for {c}osmic distance measurement{s} \citep{Eis03,BG03,HH03,SE03}. 
Since its first detection in SDSS by \cite{E05BAO}, BAO has {b}ecome essential for almost all LSS 
surveys. 
Despite the fact that $s$ is relatively large, as the Universe evolves, the accuracy {of measurements of $s$ from}
 the power spectrum (or two-point correlation function) starts to decrease rather rapidly due to  {n}on-linear 
gravitational effect{s}. 
Statistically, the information leaks into higher-order bispectra or trispectra etc., as the 
fields become more and more non-Gaussian with increased variance. 
This could be seen from the measurement of the so-called Fisher information curve, as a plateau  
starts to appear in the quasi-linear regime \citep{RH05,RH06}. 
Meanwhile, the same structure formation process moves galaxies away from their initial locations
and erase{s} the BAO feature \citep{CS08}.
Eventually, one would lose the constraining power on cosmology due to the combined effect of 
non-Gaussianity and the degraded BAO signal \citep{SE07,NHP12}.

\begin{figure*}
\centering
\includegraphics[width=1.\textwidth]{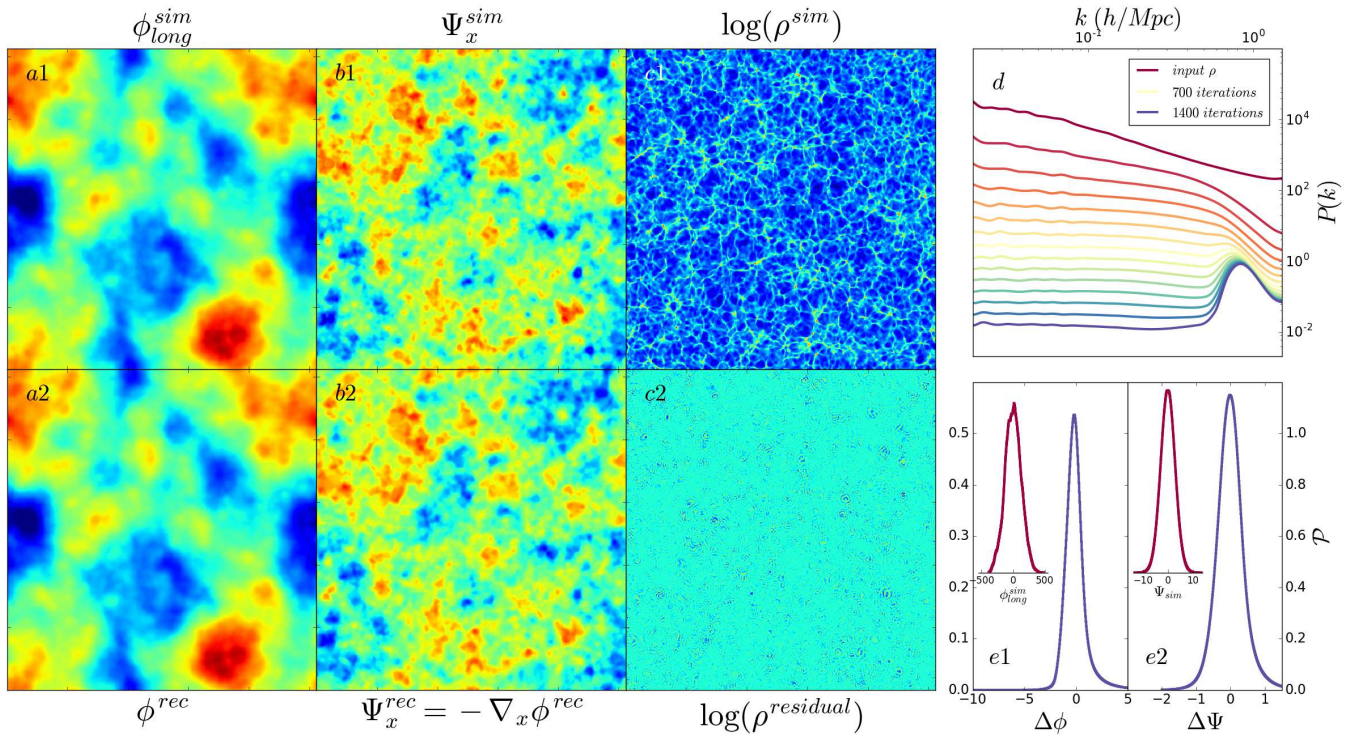}
\caption{\label{fig:snapshot} ({\it Left six panels}): {V}isual comparison of N-body simulation 
({\it top row}) and {i}sobaric reconstruction ({\it  bottom row}). From left to right, we show the
displacement potential $\phi$ ({\it panel $a1$ and $a2$}), the $x$ component of the displacement field 
({\it panel $b1$ and $b2$}), 
the {logarithm} of  {the} density field ({\it$c1$}) and the residual density map after $1400$ iterations
({\it$c2$}).  The same color scheme is applied among simulated and reconstructed field{s} for {the} {\it$a$} and 
{\it$b$} panels.
({\it Right three panels}): Statistical properties of the reconstruction, including the power spectra 
of the residual density field after many iterations ({\it$d$}){,} and the histograms of the reconstruction error  
$\Delta \phi = \phi^{sim}_{long}(\vq) - \phi^{rec}(\vxi)$ ({\it$e1$}) 
and $\Delta \Psi = \Psi^{sim}_{i} (\vq) - \Psi^{rec}_i (\vxi)$ ({\it$e2$}) at the pixel level. 
}
\end{figure*}

This situation has led to many efforts to reconstruct the BAO signature and improve the 
measurement precision of the sound horizon scale $s$ from LSS. 
For example, a logarithmic transformation or Gaussianization processes seem{s} to improve the non-Gaussianity
and bring back some information {that} leak{s} into higher order statistics \citep{Webg92,Ney11}.
On the other hand, \cite{Eis07}  {f}irst demonstrated that it is possible to sharpen 
the BAO peak in the correlation function and improve the measurement accuracy.  
This simple yet powerful algorithm {(hereafter ZA reconstruction)} has since {been} improved and applied to real measurement{s} 
\citep{PXE12,Aetal16}{, producing} stringent cosmological constraints\citep{MCX12,Aetal16}.  
Meanwhile, as the physical picture starts to clarify, more algorithms {have started} to emerge 
\citep{TZ12,EuRec15,SBZ17}.

Historically, the attempt to undo {n}on-linear structure formation dates far back before these
efforts.
\cite{Peebles89,Peebles90} first applied the principle of least action to reconstruct the trajectories 
of Local Group. This technique was further developed by other authors \citep{CG97,BFL03} 
but without any application to cosmology because of its computational cost. 
On the other hand, various approaches {have} tried to reverse the dynamical system, particularly concentrating
on the velocity potential, without worrying about any decaying modes so that observational errors will not 
overpower the real initial fluctuations \citep{NuDe92,Gram93,NC99,ME99}. 
In a method to ensure the uniqueness of the reconstruction, \cite{FMMS02} then solved the optimal mass 
transportation problem using Monge-Amp\`{e}re-Kantorovich (MAK) method \citep{BFL03,MMCS06,MS08}. 
Meanwhile, as {computational} processing power increases steadily, there are also recent attempts to 
directly sample all possible initial states using {M}arkov Chain Monte Carlo method with fast forward model{ling} 
\citep{WMJ09,JW13,WMYv13,K13}. 
Unlike most previous works,  \cite{Eis07} tried to recover the linear power spectrum instead of {the} initial 
density distribution, which is almost uniform, $\drho \ll 1$,  and therefore practically much more challenging.

From the perspective of Lagrangian dynamics, the strong non-linearity presen{t} in the density field $\drho$ 
is attributed to both the gravitational evolution of the particle movement, manifested by the 
non-linear displacement $\vPsi(\vq) = \vx - \vq$ of particles, as well as the coordinate transform between 
{the} Lagrangian and Eulerian frame{s}, where $\vx$ and $\vq$ are the Eulerian and Lagrangian position{s} respectively.  
With the assumption of the mass conservation and a uniform initial matter distribution, 
this coordinate transform could be expressed as
\begin{eqnarray}
\label{eqn:contn}
 \det \left ( \frac{ \partial x_i}{\partial q_j }\right) = \det\left ( \delta_{ij} + \partial_i \Psi_{j} \right) = 
 \frac{\bar{\rho}}{\rho }  = \frac{1}{1+\drho} .
\end{eqnarray}
Recen{t} developments suggest that this coordinate transform is the major source of various 
theoretical and practical difficulties encountered in  {l}arge-scale structure measurement{s} \citep{TsZal12,TZ12}.
Therefore, simply by avoiding this non-linear mapping, the displacement field itself would be 
much easier to handle theoretically. 
Of course, by definition, $\vPsi(\vq)$ is not an observable as it involves the unknown initial location $\vq$.
As a result, there have been efforts concentrating on the statistical extraction of $\vPsi(\vq)$, e.g. 
the maximum  likelihood method by \cite{TZ12}.

However in principle, given a non-linear density map $\drho$, and assuming only the longitudinal component of the 
displacement field,  i.e. $\vPsi=-\nabla \phi$, is cosmologically relevant, 
equation (\ref{eqn:contn}) reduces to the so-called elliptical 
Monge-Amp\`{e}re equation and becomes solvable with appropriate boundary condition{s}\footnote{
The existence and uniqueness of the solution to the Monge-Amp\`{e}re equation is out the scope of this paper. 
Theorems {were} proven by \cite{B57,A58} for the generalized solution with Dirichlet boundary condition{s} and
 \cite{LTU86}  for the Neumann problem. One could also see e.g. \cite{GW08} for more detailed discussion.
}. 
Of course, this assumption only holds if the input map $\drho$ is generated from particl{e} movement without 
any stream-crossing or rotation. 
In realit{y,} these transverse components do emerge, but only at smaller scales \citep{Chan14}. 
Therefore the {induced} reconstruction error in the BAO regime would be negligible. 

Recently, \cite{ZHM16} first propose{d} the {\it potential isobaric reconstruction}  algorithm 
to reverse equation (\ref{eqn:contn}) and solve for the displacement potential $\phi$, and demonstrated {that the resulting reconstruction of displacement potential is highly correlated with the} initial density field{.} 
\cite{Pan16} {further} showed that the Fisher information of the reconstructed field {is} significantly 
increased especially {in the} quasi-linear regime, an{d} \cite{YZP17} {performed a similar study} for {the} halo density field. 
Shortly after {these papers were released}, \cite{SBZ17} also proposed a particle-based iterative 
reconstruction algorithm that performs similarly.

In this paper, we apply the same algorithm and particularly concentrate o{n}
cosmological constraint{s} from measurin{g} baryonic acoustic oscillation{s}.   
In section 2., we first review this algorithm and demonstrate {its} ability to recover the nonlinear
displacement field. 
We focus on the BAO signature in section 3., and show {how the algorithm can improve the accuracy of} the recovered wiggles and {thereby improve} cosmological measurement{s}.  Finally, we conclude and discuss in section 4.  
Throughout this paper, the N-body simulations were run with numerical code CUBE, with the box size of
 $1024~ \Mpc/h$, and $512^3$ particles. We adopt the Planck 2015 results \citep{Planck2015cos}
as our fiducial cosmological parameters.

\section{The Potential Isobaric Reconstruction}

As a non-linear partial differential equation, the numerical solution to equation (\ref{eqn:contn}) is 
nontrivial{. Various numerical strategies, e.g. \cite{MAeqZPF10,LFOX14}, exist in the literature, but} one cavea{t} of directly applying these algorithms in cosmology is the frame mismatch between the 
Lagrangian gradient of the displacement $\partial_i \Psi_{j} = \partial \Psi_i/ \partial q_j$ and 
the Eulerian density  field $\rho(\vx)$ 
\footnote{It has been demonstrated that with the assumption of no stream-crossing, 
ther{e} exists {a} potential field $\theta(\vx)$ that {satisfies} $\vq=\vx-\nabla_x \theta(\vx)$  
\citep{Arn79,FMMS02}, {and} therefore equation (\ref{eqn:contn}) is indeed well-defined and solvable with 
standard algorithm{s}.}.
In this paper, we follow the multi-grid moving mesh algorithm proposed by \cite{Pen95,Pen98} to iteratively 
solve the differential form of this mass conservation equation in a moving curvilinear coordinate system, 
which we name as the {\it potential isobaric coordinate}.

\begin{figure}
\centering
\includegraphics[width=0.5\textwidth]{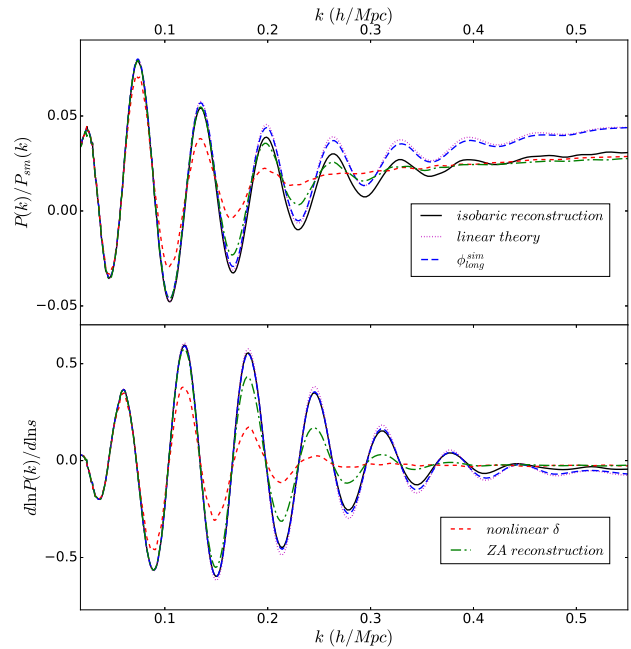}
\caption{  \label{fig:BAO}
Comparison of BAO ({\it upper}) and the dependence of the power spectra with 
respect to the sound horizon $s$ scale ({\it lower}) at redshift $z=0$. 
To eliminate the sample variance, we run a few N-body simulations and reconstructions with different 
transfer functions (with different assumptions of BAO) but the same initial random seeds. 
For ZA reconstruction \citep{Eis07}, we assume a Gaussian smoothing with $R=15 ~\Mpc/h$.}
\end{figure}

To present our method, we first take the differentiation of equation (\ref{eqn:contn}) with 
respect to some time-like iteration parameter $\tau$, which gives the conservation equation
\begin{eqnarray}
\label{eqn:MAElin}
d_{\tau} \left[ \rho\left(\vx\left(\vq \right) \right)   \det\left( \frac{\partial x_i}{ \partial q_j }\right ) 
\left(\vq \right)  \right] = 0  
\end{eqnarray}
of a unit Lagrangian volume. 
For given Eulerian cell, it further simplifies to  $d_{\tau} \rho = 0$.   
Notice that both equations (\ref{eqn:contn}) and (\ref{eqn:MAElin}) are expressed locally, {and} therefore
do not explain how the conservation would be maintained as long as the density and volume of the cell 
vary consistently. 
But from the perspective of the field, equation (\ref{eqn:MAElin}) is nothing but the continuity equation with 
respect to $\tau$, which is the equation we will solve.
To trace the deformation of the fluid elements, we work in a moving numerical grid with curvilinear coordinate
 $\vxi = \{\xi^{\mu} \}  = \{\xi^1, \xi^2, \xi^3 \}$. To determine the physical position of the lattice, one needs to
specify the coordinate transform relation $x^i(\xi^{\mu})$. 
Following \cite{Pen95,Pen98}, we borrow the notation fro{m} general relativity to describe a general 
curvilinear coordinate system. 
In this paper, we use Greek indices for curvilinear grid coordinate{s}, and Latin indices for Cartesian coordinate{s}.
The coordinate transform is described by the triad $e^i_{\mu} = \partial x^i / \partial \xi^{\mu}$. 
The flat metric in Cartesian coordinate{s} $x^i$ is simply the Kronecker delta $g_{ij} = \delta_{ij}$, while the 
curvilinear metric is given by $g_{\mu\nu} =  e^i_{\mu} e^j_{\nu} \delta_{ij}$, both of which could be used to raise 
or lower the indices. 
Particularly we adopt the following ansatz at each iteration step    
\begin{eqnarray}
\label{eqn:curv_grid}
x^i (\xi^{\mu}) =  \xi^{\mu} \delta^i_{\mu} + d x^i 
= \xi^{\mu} \delta^i_{\mu} + \frac{\partial (d \phi ) }{\partial \xi^{\nu}} \delta^{i\nu} ,
\end{eqnarray}
where  $\delta^i_{\mu}$ is the Kronecker delta function, and we use Einstein summation convention. 
Here $d\phi(\xi^{\mu})$ is the increment of the displacement potential{;} since we neglect the transverse component,
this ansatz will help to minimize the mesh twisting.

As demonstrated by \cite{Pen98}, in the isobaric frame, in which the mass per grid cell is kept 
constant, the continuity equation could be expressed as a linear elliptic equation
\begin{eqnarray}
\label{eqn:poisson_curv}
\partial_{\mu} \left( \rho \sqrt{g}  e^{\mu}_i \delta^{i\nu} \partial_{\nu} \left(d_{\tau} \phi \right) \right) =   
 \partial_{\mu} \left( \rho \sqrt{g} e^{\mu}_i \left(d_{\tau} x^i\right) \right)
= d_{\tau} \rho, 
\end{eqnarray}
where the dreibein $e^{\mu}_i$ is the inverse of the triad, or the Jacobian matrix, 
$e^i_{\mu} = \partial x^i/\partial \xi^{\mu}$, the volume element  $\sqrt{g} = \det(e^i_{\mu})$.
For given non-linear density map $\rho(\vx)$, we would like to reverse the frame deformation, and 
solve for the increment of potential $d\phi$, where $d\phi = d_{\tau} \phi d\tau=d_{\tau} \phi$
with the assumption that $d\tau = 1$ since we are free to parametrize $\tau$. 
As shown in \cite{ZHM16}, this is done by substituting $d\rho$ in equation (\ref{eqn:poisson_curv}) 
with the negative non-linear density contrast $-\Delta \rho =  \bar{\rho} -\sqrt{g} \rho$  together with 
a spatial smoothing $S$ to ensure the resultant $d\phi$ is small enough, i.e.
\begin{eqnarray}
\label{eqn:dphi_curv}
\partial_{\mu} \left(  \rho \sqrt{g} e^{\mu}_i \delta^{i\nu} \partial_{\nu} (d \phi) \right) =
 S ( -\Delta \rho + C + E). 
\end{eqnarray}
Here $C$ and $E$ are the compression and expansion limiters for improving the numerical stability 
\citep{Pen98}.
Consequently, the triad $e^i_{\mu}$ {is} alway{s} kept positive definite. 
Then after moving the grid by equation (\ref{eqn:curv_grid}), we estimate the density increment 
$d\rho$ from equation (\ref{eqn:poisson_curv}) and then repeat equation 
(\ref{eqn:dphi_curv}) with re-evaluated 
density contrast $\rho^{(i+1)} = \rho^{(i)} + d\rho^{(i)}$.

Eventually, the reconstructed displacement potential is simply a summation of all iterations, 
\begin{eqnarray}
\phi^{rec} (\vxi) = \sum_i^{iters}  d\phi^{(i)} \left(\vxi^{(i)}  \left (\vxi \right) \right ) .
\end{eqnarray}
Here we have explicitly denoted the {\it one-to-one} mapping between the intermediate and final 
coordinate  $\vxi^{(i)} \left(  \vxi \right) $, which is only possible because of the positive definiteness of the 
transform matrix $e^i_{\mu}$ we achieved in solving equation (\ref{eqn:dphi_curv}).
As a result, our final product $\phi^{rec}$ is then charted in a frame $\vxi$ that is almost identical
to the theoretically defined Lagrangian coordinate $\vq$, which means we have 
$\phi^{rec}(\vxi) \approx \phi^{sim}_{long} (\vq)$ at the {\it pixel level}.

In the first column of Figure  \ref{fig:snapshot}., i.e. panel $a1$ and $a2$ (with the same colorbar), 
we compare the simulation displacement potential  $\phi^{sim}_{long}$ ({\it panel $a2$})
with the reconstructed potential $\phi^{rec}$  ({\it panel $a1$}).
Here, $\phi^{sim}_{long}$ is calculated simply by taking the longitudinal component of the 
real displacement. 
The visual difference is not very obvious even when it is compared with the full displacement vector field, 
i.e.  $\Psi^{sim}_x$ ({\it panel $b1$}) vs. $\Psi^{rec}_x=-\nabla_x \phi^{rec}$ ({\it panel $b2$}), when 
transverse contribution starts to emerge after the shell-crossing. 
As shown in the histogram of $\Delta \phi = \phi^{sim}_{long}(\vq) - \phi^{rec}(\vxi)$  ({\it panel $e1$}), 
the reconstruction error is within a percent for $\phi^{rec}$, 
and is around ten-percent for {the} displacement $\vPsi$ itself ({\it panel $e2$}).

In {panels} $c1$ and $c2$, we also visually compare the input density map and its residual after 
$1400$ iterations. As shown, the non-vanishing residuals are basically around high density region{s}, where the 
displacement field $\vPsi$ would no longer be captured by longitudinal component $-\nabla \phi$ alone anymore.  
From the power spectra of the residual map in panel $d$, this starts to occur around the comoving
Fourier scale $k \gtrsim 0.6 h/\Mpc$ and peaks around $0.8 h/\Mpc$.
Even though the large scale power has been reduced to about six orders of magnitude smaller than the input,  
the small-scale power is still around unity.

\section{Measuring the Baryonic Acoustic Oscillation}
We are particularly interested in applying this isobaric reconstruction algorithm to detect
the BAO signature in the power spectrum.
As demonstrated by various authors, e.g. \cite{CS08}, the sharp BAO 
features from the linear $P_{\rm ini}(k)$ {are} smeared out by the bulk flow of the galaxies.
This could be seen by simply expressing the nonlinear density $\drho$ from the continuity equation 
(\ref{eqn:contn}) in Fourier space, and then taking the power spectrum \citep{CS06a,CS08}
\begin{eqnarray}
P_{\rm nl}(k) &=&  \int \frac{d^3 r}{(2\pi)^3} e^{i \vk \cdot \vrr}  
\left [  \langle e^{i \vk \cdot \Delta \vPsi } \rangle  -1  \right]  \nonumber\\
&=& G^2(k) P_{\rm ini} (k) + P_{\rm mc}(k). 
\end{eqnarray}
Here $\vrr = \vq - \vq^{\pri}$,  $\Delta \vPsi = \vPsi(\vq) - \vPsi(\vq^{\pri}) $, 
$G(k)$ is the so-called non-linear propagator  $G^2(k) \approx \langle e^{i \vk \cdot \Delta \vPsi} \rangle  
\approx  \exp(-k^2 \sigma^2_v)$, 
$P_{\rm mc}(k)$ is the model-coupling power spectrum and $\sigma_v^2$ is the variance of the velocity.
Since $P_{\rm mc}(k)$ dominates at large $k$ and is quite smooth, the BAO feature which could be 
defined as dividing $P_{{\rm nl}}(k)$ by some no-wiggle template $P_{\rm sm}(k)$, i.e. 
\begin{eqnarray}
{\rm Wig}(k)=\frac{P_{\rm nl}(k)}{P_{\rm sm} (k)} -1 \approx \exp(-k^2 \sigma^2_v ) ~ {\rm Wig}_{\rm ini} (k),
\end{eqnarray}
would be smeared out by the bulk flow $\Delta \vPsi$. 
Clearly, the reconstructed displacement field will no longer be affected by this smearing.

\begin{figure}
\centering
\includegraphics[width=0.5\textwidth]{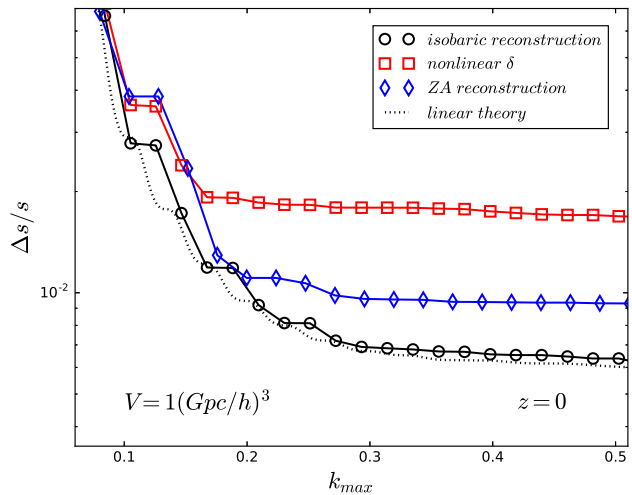}
\caption{ \label{fig:dserr}
The fractional error of measuring the sound horizon scale $\Delta s/s$ as a function of maximum $k$
for a noiseless survey of volume 
$1 ({\rm Gpc}/h)^3$ at redshift $z=0$. The measurement precision after the isobaric reconstruction is
very close to the linear theory, which assumes both linear power spectra and Gaussian 
covariance matrix (equation \ref{eqn:cov_gau}).  
The error after the reconstruction is roughly $2.7$ times smaller than original density field, 
compared to a factor of $\sim 1.8$ smaller for the ZA reconstruction method. 
}
\end{figure}

To examine the reconstructed BAO signature, we run a few N-body simulations with different 
transfer functions (with/without BAO, and with different sound horizon scale $s$)  but the same initial random seeds. 
This enables us to reveal the signal with little numerical efforts but is free from the sample variance.
As shown in the upper panel of Figure \ref{fig:BAO}., at redshift $z=0$, BAO signature becomes basically
invisible after the fourth pea{k} ({\it red-dashed line}), whereas the non-linear displacement potential 
$\phi^{sim}_{long}$ ({\it blue-dashed line}) is almost identical to the initial density field ({\it red-dotted line}) in this
$k$ range. 
The BAO of the reconstructed field $\phi^{rec}$ ({\it black-solid line}), are recovered with very high 
fidelity, although there seems to exist a percent-level change of broad band power. 
With an appropriate template, the sound horizon scale $s$ would be recovered at a level that would be very close 
to the linear theory. This is clear from the numerical derivative $d\ln P(k)/d\ln s$ in the lower panel of Figure 
\ref{fig:BAO}, which is derived from the reconstruction of simulations with slightly different $s$.
As a comparison, we also display the performance of ZA reconstruction method \citep{Eis07} assuming 
a Gaussian smoothing length $15 \Mpc/h$.

To evaluate the impact on the constraints of the cosmic distance, we estimate the Fisher information 
on the sound horizon scale $s$ for a noiseless survey. 
By definition, the Fisher matrix is 
\begin{eqnarray}
F_{\Delta \ln s} = \frac{\partial \ln P(k_i)}{\partial \ln s} C^{-1}_{ij}  \frac{\partial \ln P(k_j)}{\partial \ln s} .
\end{eqnarray}
Here $C_{ij}$ is the normalized covariance matrix 
\begin{eqnarray}
C_{ij} = \frac{1}{n_{s}-1} \sum_{m}^{n_{s}} 
\left ( \frac{ P_m(\vk_i )}{\langle P(\vk_i) \rangle} -1  \right) 
\left ( \frac{ P_m(\vk_j )}{\langle P(\vk_j) \rangle} -1  \right), 
\end{eqnarray}
which is estimated from $n_s = 200$ N-body simulations.
We then rescaled the large-scale component ($k < 0.1$) of matrix $C_{ij}$ to 
coincide with a Gaussian covariance matrix $C^g_{ij}$ 
\begin{eqnarray}
\label{eqn:cov_gau}
C^g_{ij} = \frac{2}{N_k} \delta^K_{ij}  
= \frac{4 \pi^2 }{ k_i^2 \Delta k_i V_{s} }  \delta^K_{ij}  
\end{eqnarray}
assuming a survey volume of $V_{s}=1~({\rm Gpc}/h)^3$, {w}here $N_k$ is the number of $k$ modes 
{in} such {a} survey.

In Figure \ref{fig:dserr}., we show that the fractional error on the sound horizon $\Delta s/s$ measured
from our reconstructed field is roughly $2.7$ times smaller than from the non-linear density field itself 
({\it red dashed line}). 
This is very close to the accuracy one could ever achieve with the linear 
BAO ({\it black-dotted line}) feature and Gaussian covariance matrix.  
As a comparison, the same number is roughly $\sim 1.8$ for the ZA reconstruction.

\section{Discussion}
As the method of circumventing the strongest non-linearity in the structure formation process, we 
believe our reconstruction algorithm will significantly affect the way we measure large-scale structure 
in the future. 
However, in this paper, we did not take into account various complications, e.g. {r}edshift space 
distortion{s} (RSD) {and} biasing. 
In {r}edshift space, the reconstructed field is simply a combination
of the original displacement with the peculiar velocity{,}
$\vPsi_{rsd} = \vPsi + \vv/\mH$. With some processing, e.g.~{a} Wiener filter, one could then
recover $\vPsi$.
The bias, on the other hand, is probably  more challenging.
Without any preprocessing, one could perform the reconstruction for the biased density field itself,  
but the output $\phi^{rec}_{b}$ will then contain the bias information, and {will be} mapped on a grid other 
than $\vq$. 
Alternatively, one could simply fix the bias parameters from the unconstructed density field, e.g. via HOD 
fitting or simply linear bias \citep{PXE12}.

\acknowledgments
XW would like to thank Simon Foreman for proofreading the manuscript.

\newcommand{\appsection}[1]{\let\oldthesection\thesection
\renewcommand{\thesection}{\oldthesection}
\section{#1}\let\thesection\oldthesection}


\label{lastpage}

\end{document}